\pdfoutput=1
\documentclass[12pt,a4paper]{article}
\usepackage{ifthen} 
\newboolean{pdflatex}
\setboolean{pdflatex}{true} 

\newboolean{articletitles}
\setboolean{articletitles}{true} 

\newboolean{uprightparticles}
\setboolean{uprightparticles}{false} 

\newboolean{inbibliography}
\setboolean{inbibliography}{false} 

\usepackage{microtype}
\usepackage{lineno} 
\usepackage{xspace} 
\usepackage{caption}

\usepackage{graphicx}  
\usepackage{color}
\usepackage{colortbl}
\graphicspath{{./}} 

\usepackage{amsmath} 
\usepackage{amssymb}
\usepackage{amsfonts}
\usepackage{upgreek} 

\usepackage{accents}

\newcommand*\patchAmsMathEnvironmentForLineno[1]{%
        \expandafter\let\csname old#1\expandafter\endcsname\csname #1\endcsname
        \expandafter\let\csname oldend#1\expandafter\endcsname\csname
        end#1\endcsname
        \renewenvironment{#1}%
        {\linenomath\csname old#1\endcsname}%
        {\csname oldend#1\endcsname\endlinenomath}%
}
\newcommand*\patchBothAmsMathEnvironmentsForLineno[1]{%
        \patchAmsMathEnvironmentForLineno{#1}%
        \patchAmsMathEnvironmentForLineno{#1*}%
}
\AtBeginDocument{%
        \patchBothAmsMathEnvironmentsForLineno{equation}%
        \patchBothAmsMathEnvironmentsForLineno{align}%
        \patchBothAmsMathEnvironmentsForLineno{flalign}%
        \patchBothAmsMathEnvironmentsForLineno{alignat}%
        \patchBothAmsMathEnvironmentsForLineno{gather}%
        \patchBothAmsMathEnvironmentsForLineno{multline}%
        \patchBothAmsMathEnvironmentsForLineno{eqnarray}%
}

\usepackage{hyperref}    
\usepackage[all]{hypcap}

\def\lhcb {\mbox{LHCb}\xspace}

\def\MagUp {\mbox{\em Mag\kern -0.05em Up}\xspace}

\ifthenelse{\boolean{uprightparticles}}%
{

 \def\Ppi         {\ensuremath{\uppi}\xspace}

 \def\Ppsi        {\ensuremath{\uppsi}\xspace}

 \def\PDelta      {\ensuremath{\Delta}\xspace}                 
 \def\PXi      {\ensuremath{\Xi}\xspace}                 
 \def\PLambda      {\ensuremath{\Lambda}\xspace}                 
 \def\PSigma      {\ensuremath{\Sigma}\xspace}                 
 \def\POmega      {\ensuremath{\Omega}\xspace}                 
 \def\PUpsilon      {\ensuremath{\Upsilon}\xspace}

 \def\PB      {\ensuremath{\mathrm{B}}\xspace}                 
                  
 \def\PD      {\ensuremath{\mathrm{D}}\xspace}

 \def\PJ      {\ensuremath{\mathrm{J}}\xspace}                 
 \def\PK      {\ensuremath{\mathrm{K}}\xspace}

 \def\Pb      {\ensuremath{\mathrm{b}}\xspace}                 
 \def\Pc      {\ensuremath{\mathrm{c}}\xspace}

 \def\Pi      {\ensuremath{\mathrm{i}}\xspace}

 \def\Ps      {\ensuremath{\mathrm{s}}\xspace}                 
 \def\Pt      {\ensuremath{\mathrm{t}}\xspace}

}
{

 \def\Ppi         {\ensuremath{\pi}\xspace}

 \def\Ppsi        {\ensuremath{\psi}\xspace}                 
                  
 \mathchardef\PDelta="7101
 \mathchardef\PXi="7104
 \mathchardef\PLambda="7103
 \mathchardef\PSigma="7106
 \mathchardef\POmega="710A
 \mathchardef\PUpsilon="7107
                  
 \def\PB      {\ensuremath{B}\xspace}                 
                  
 \def\PD      {\ensuremath{D}\xspace}

 \def\PJ      {\ensuremath{J}\xspace}                 
 \def\PK      {\ensuremath{K}\xspace}

 \def\Pb      {\ensuremath{b}\xspace}                 
 \def\Pc      {\ensuremath{c}\xspace}

 \def\Pi      {\ensuremath{i}\xspace}

 \def\Ps      {\ensuremath{s}\xspace}                 
 \def\Pt      {\ensuremath{t}\xspace}

}

\makeatletter
\ifcase \@ptsize \relax
  \newcommand{\miniscule}{\@setfontsize\miniscule{4}{5}}
\or
  \newcommand{\miniscule}{\@setfontsize\miniscule{5}{6}}
\or
  \newcommand{\miniscule}{\@setfontsize\miniscule{5}{6}}
\fi
\makeatother

\DeclareRobustCommand{\optbar}[1]{\shortstack{{\miniscule (\rule[.5ex]{1.25em}{.18mm})}
  \\ [-.7ex] $#1$}}

\def\squark    {{\ensuremath{\Ps}}\xspace}

\def\cquark    {{\ensuremath{\Pc}}\xspace}
\def\cquarkbar {{\ensuremath{\overline \cquark}}\xspace}
\def\ccbar     {{\ensuremath{\cquark\cquarkbar}}\xspace}
\def\bquark    {{\ensuremath{\Pb}}\xspace}

\def\tquark    {{\ensuremath{\Pt}}\xspace}

\def\pion   {{\ensuremath{\Ppi}}\xspace}

\def\pip    {{\ensuremath{\pion^+}}\xspace}
\def\pim    {{\ensuremath{\pion^-}}\xspace}

\def\kaon    {{\ensuremath{\PK}}\xspace}
  \def\Kbar    {{\kern 0.2em\overline{\kern -0.2em \PK}{}}\xspace}

\def\KorKbar    {\kern 0.18em\optbar{\kern -0.18em K}{}\xspace}

\def\Kp      {{\ensuremath{\kaon^+}}\xspace}
\def\Km      {{\ensuremath{\kaon^-}}\xspace}

  \def\Dbar    {{\kern 0.2em\overline{\kern -0.2em \PD}{}}\xspace}

\def\D       {{\ensuremath{\PD}}\xspace}
\def\DorDs       {{\ensuremath{\PD_{(\squark)}}}\xspace}

\def\DorDbar    {\kern 0.18em\optbar{\kern -0.18em D}{}\xspace}
\def\Dz      {{\ensuremath{\D^0}}\xspace}

\def\Dp      {{\ensuremath{\D^+}}\xspace}
\def\Dm      {{\ensuremath{\D^-}}\xspace}

\def\Dstarp  {{\ensuremath{\D^{*+}}}\xspace}

\def\Dsp     {{\ensuremath{\D^+_\squark}}\xspace}
\def\Dsm     {{\ensuremath{\D^-_\squark}}\xspace}

\def\Dssp    {{\ensuremath{\D^{*+}_\squark}}\xspace}
\def\Dssm    {{\ensuremath{\D^{*-}_\squark}}\xspace}

\def\B       {{\ensuremath{\PB}}\xspace}
\def\BorBs       {{\ensuremath{\PB_{(\squark)}}}\xspace}
\def\Bbar    {{\ensuremath{\kern 0.18em\overline{\kern -0.18em \PB}{}}}\xspace}

\def\BorBbar    {\kern 0.18em\optbar{\kern -0.18em B}{}\xspace}
\def\Bz      {{\ensuremath{\B^0}}\xspace}

\def\Bub     {{\ensuremath{\B^-}}\xspace}

\def\Bm      {{\ensuremath{\Bub}}\xspace}

\def\Bd      {{\ensuremath{\B^0}}\xspace}
\def\Bs      {{\ensuremath{\B^0_\squark}}\xspace}
\def\Bsb     {{\ensuremath{\Bbar{}^0_\squark}}\xspace}

\def\jpsi     {{\ensuremath{{\PJ\mskip -3mu/\mskip -2mu\Ppsi\mskip 2mu}}}\xspace}

  \def\Y#1S{\ensuremath{\PUpsilon{(#1S)}}\xspace}

\def\Lbar        {{\ensuremath{\kern 0.1em\overline{\kern -0.1em\PLambda}}}\xspace}
\def\LorLbar    {\kern 0.18em\optbar{\kern -0.18em \PLambda}{}\xspace}

\newcommand{\decay}[2]{\ensuremath{#1\!\to #2}\xspace}         

\def\to                 {\ensuremath{\rightarrow}\xspace}

\def\CP                {{\ensuremath{C\!P}}\xspace}

\def\Vcs  {{\ensuremath{V_{\cquark\squark}}}\xspace}
\def\Vts  {{\ensuremath{V_{\tquark\squark}}}\xspace}

\def\Vcb  {{\ensuremath{V_{\cquark\bquark}}}\xspace}
\def\Vtb  {{\ensuremath{V_{\tquark\bquark}}}\xspace}

\newcommand{\dms}{{\ensuremath{\Delta m_{\squark}}}\xspace}

\newcommand{\DGs}{{\ensuremath{\Delta\Gamma_{\squark}}}\xspace}

\newcommand{\Gs}{{\ensuremath{\Gamma_{\squark}}}\xspace}

\newcommand{\mL}{{\ensuremath{m_{\rm L}}}\xspace}
\newcommand{\mH}{{\ensuremath{m_{\rm H}}}\xspace}
\newcommand{\GL}{{\ensuremath{\Gamma_{\rm L}}}\xspace}
\newcommand{\GH}{{\ensuremath{\Gamma_{\rm H}}}\xspace}

\newcommand{\phis}{{\ensuremath{\phi_{\squark}}}\xspace}
\newcommand{\betas}{{\ensuremath{\beta_{\squark}}}\xspace}

\newcommand{\fancybar}{\scalebox{.4}{(}\raisebox{-1.7pt}{--}\scalebox{.4}{)}}
\newcommand{\brabar}[1]{\accentset{\fancybar}{#1}}
\newcommand{\mistag}{\ensuremath{\omega}\xspace}

\newcommand{\etag}{{\ensuremath{\varepsilon_{\rm tag}}}\xspace}

\newcommand{\effD}{{\ensuremath{\etag D^2}}\xspace}

\def\DsDs {\Dsp\Dsm}
\def\DpDs {\Dm\Dsp}

\def\BsToDsDs {\decay{\Bsb}{\Dsp\Dsm}}
\def\BzToDsD {\decay{\Bz}{\Dm\Dsp}}
\def\BsToDsD {\decay{\Bsb}{\Dsp\Dm}}
\def\BToJpsiK{\decay{\Bm}{\jpsi\Km}}

\def\BsToDsstDsst{\decay{\Bsb}{\Dssp\Dssm}}

\def\BsToDsPi  {\decay{\Bsb}{\Dsp\pim}}
\def\BsToJPsiPhi  {\decay{\Bsb}{\jpsi\phi}}

\def\BsToJPsiPiPi  {\decay{\Bsb}{\jpsi\pip\pim}}
\def\BsToJPsiKK  {\decay{\Bsb}{\jpsi\Kp\Km}}

\def\AT#1     {\ensuremath{A_{\mathrm{T}}^{#1}}\xspace}           

\def\C#1      {\ensuremath{\mathcal{C}_{#1}}\xspace}                       
\def\Cp#1     {\ensuremath{\mathcal{C}_{#1}^{'}}\xspace}                    
\def\Ceff#1   {\ensuremath{\mathcal{C}_{#1}^{\mathrm{(eff)}}}\xspace}        
\def\Cpeff#1  {\ensuremath{\mathcal{C}_{#1}^{'\mathrm{(eff)}}}\xspace}       
\def\Ope#1    {\ensuremath{\mathcal{O}_{#1}}\xspace}                       
\def\Opep#1   {\ensuremath{\mathcal{O}_{#1}^{'}}\xspace}                    

\newcommand{\tev}{\ifthenelse{\boolean{inbibliography}}{\ensuremath{~T\kern -0.05em eV}\xspace}{\ensuremath{\mathrm{\,Te\kern -0.1em V}}}\xspace}
\newcommand{\gev}{\ensuremath{\mathrm{\,Ge\kern -0.1em V}}\xspace}
\newcommand{\mev}{\ensuremath{\mathrm{\,Me\kern -0.1em V}}\xspace}
\newcommand{\kev}{\ensuremath{\mathrm{\,ke\kern -0.1em V}}\xspace}
\newcommand{\ev}{\ensuremath{\mathrm{\,e\kern -0.1em V}}\xspace}
\newcommand{\gevc}{\ensuremath{{\mathrm{\,Ge\kern -0.1em V\!/}c}}\xspace}
\newcommand{\mevc}{\ensuremath{{\mathrm{\,Me\kern -0.1em V\!/}c}}\xspace}
\newcommand{\gevcc}{\ensuremath{{\mathrm{\,Ge\kern -0.1em V\!/}c^2}}\xspace}
\newcommand{\gevgevcccc}{\ensuremath{{\mathrm{\,Ge\kern -0.1em V^2\!/}c^4}}\xspace}
\newcommand{\mevcc}{\ensuremath{{\mathrm{\,Me\kern -0.1em V\!/}c^2}}\xspace}

\def\invfb   {\ensuremath{\mbox{\,fb}^{-1}}\xspace}

\def\ps   {\ensuremath{{\rm \,ps}}\xspace}
\def\fs   {\ensuremath{\rm \,fs}\xspace}

\def\invps{\ensuremath{{\rm \,ps^{-1}}}\xspace}

\newcommand{\stat}{\ensuremath{\mathrm{\,(stat)}}\xspace}
\newcommand{\syst}{\ensuremath{\mathrm{\,(syst)}}\xspace}

\def\gsim{{~\raise.15em\hbox{$>$}\kern-.85em
          \lower.35em\hbox{$\sim$}~}\xspace}
\def\lsim{{~\raise.15em\hbox{$<$}\kern-.85em
          \lower.35em\hbox{$\sim$}~}\xspace}

\def\sPlot{\mbox{\em sPlot}\xspace}

\def\mrad{\ensuremath{\rm \,mrad}\xspace}
\def\rad{\ensuremath{\rm \,rad}\xspace}

\def\tell1  {TELL1\xspace}
\def\ukl1   {UKL1\xspace}

\newcommand{\ie}{\mbox{\itshape i.e.}\xspace}

\usepackage{cite} 
\usepackage{mciteplus}

\usepackage{longtable} 

\begin{document}

\renewcommand{\thefootnote}{\fnsymbol{footnote}}
\setcounter{footnote}{1}

\begin{titlepage}
\pagenumbering{roman}

\vspace*{-1.5cm}
\centerline{\large EUROPEAN ORGANIZATION FOR NUCLEAR RESEARCH (CERN)}
\vspace*{1.5cm}
\hspace*{-0.5cm}
\begin{tabular*}{\linewidth}{lc@{\extracolsep{\fill}}r}
\ifthenelse{\boolean{pdflatex}}
{\vspace*{-2.7cm}\mbox{\!\!\!\includegraphics[width=.14\textwidth]{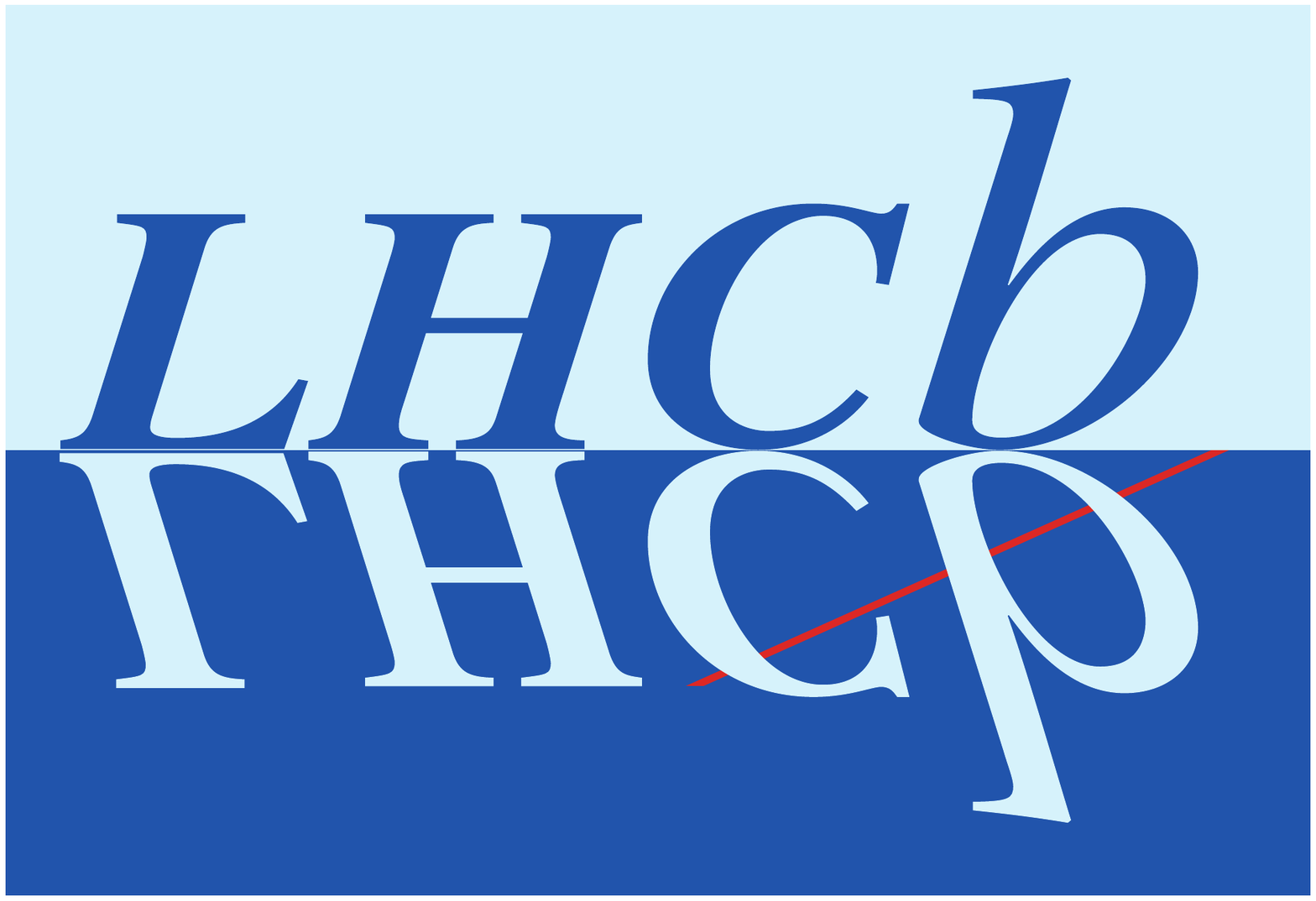}} & &}%
{\vspace*{-1.2cm}\mbox{\!\!\!\includegraphics[width=.12\textwidth]{lhcb-logo.eps}} & &}%
\\
 & & CERN-PH-EP-2014-223 \\  
 & & LHCb-PAPER-2014-051 \\  
 & & September 16, 2014 \\ 
 & & \\
\end{tabular*}

\vspace*{4.0cm}

{\bf\boldmath\huge
\begin{center}
  Measurement of the \CP-violating phase \phis in \BsToDsDs decays
\end{center}
}

\vspace*{2.0cm}

\begin{center}
The LHCb collaboration\footnote{Authors are listed at the end of this Letter.}
\end{center}

\vspace{\fill}

\begin{abstract}
  \noindent
  We present a measurement of the \CP-violating weak mixing phase \phis using the decay \BsToDsDs in a data sample corresponding to 3.0\invfb of integrated luminosity collected with the \lhcb detector in $pp$ collisions at centre-of-mass energies of 7 and 8\tev. An analysis of the time evolution of the system, which does not use the constraint $|\lambda|=1$ to allow for the presence of \CP violation in decay, yields $\phis = 0.02 \pm 0.17\stat \pm 0.02\syst\rad$, $|\lambda| = 0.91~^{+0.18}_{-0.15}\stat\pm0.02\syst$.  This result is consistent with the Standard Model expectation.
\end{abstract}

\vspace*{2.0cm}

\begin{center}
 Submitted to Phys.~Rev.~Lett.
\end{center}

\vspace{\fill}

{\footnotesize
\centerline{\copyright~CERN on behalf of the \lhcb collaboration, license \href{http://creativecommons.org/licenses/by/4.0/}{CC-BY-4.0}.}}
\vspace*{2mm}

\end{titlepage}

\newpage
\setcounter{page}{2}
\mbox{~}

\cleardoublepage

\renewcommand{\thefootnote}{\arabic{footnote}}
\setcounter{footnote}{0}

\pagestyle{plain} 
\setcounter{page}{1}
\pagenumbering{arabic}

The \CP-violating weak mixing phase $\phis$ can be measured in the interference between mixing and decay of \Bsb mesons to \CP eigenstates that proceeds via the $\bquark\to\ccbar\squark$ transition, and is predicted to be small in the Standard Model (SM): $\phi^{\text{SM}}_{s}\approx -2\betas \equiv -2\arg\left ( -\frac{\Vts\Vtb^{*}}{\Vcs\Vcb^{*}}\right ) = -36.3^{+1.6}_{-1.5}\mrad$~\cite{Charles:2011va}. Measurements of $\phis$ are sensitive to the effects of potential non-SM particles contributing to the \Bs-\Bsb mixing amplitude.
Several measurements of $\phis$ have been made with the decay mode \BsToJPsiPhi, with the first results showing tension with the SM expectation~\cite{Aaltonen:2007he,Abazov:2008af}. Since then, more recent measurements of $\phis$ have found values consistent with the SM prediction in \BsToJPsiKK and \BsToJPsiPiPi decays~\cite{LHCb-PAPER-2013-002,LHCb-PAPER-2014-019,Abazov:2011ry,Aad:2012kba, Aaltonen:2012ie}. The world average value determined prior to the publication of Ref.~\cite{LHCb-PAPER-2014-019} is $\phis=0\pm70\mrad$~\cite{HFAG}. 

Precise measurements of $\phis$ are complicated by the presence of loop (penguin) diagrams, which could have an appreciable effect~\cite{Faller:2008gt}. It is therefore important to measure $\phis$ in additional decay modes where penguin amplitudes may differ~\cite{Fleischer:2007zn}. Additionally, in the \BsToJPsiPhi channel, where a spin-0 meson decays to two spin-1 mesons, an angular analysis is required to disentangle statistically the \CP-even and \CP-odd components. The decay \BsToDsDs is also a $\bquark\to\ccbar\squark$ transition with which $\phis$ can be measured~\cite{Dunietz:2000cr}, with the advantage that the \DsDs final state is \CP-even, and does not require angular analysis.

In this Letter, we present the first measurement of \phis in \BsToDsDs decays using an integrated luminosity of 3.0\invfb, obtained from $pp$ collisions collected by the \lhcb detector. One third of the data were collected at a centre-of-mass energy of 7\tev, and the remainder at 8\tev. 
We perform a fit to the time evolution of the \Bsb-\Bs system in order to extract \phis.

\lhcb is a single-arm forward spectrometer at the LHC designed for the study of particles containing \bquark or \cquark quarks in the \mbox{pseudorapidity} range 2 to 5~\cite{Alves:2008zz}. Events are selected by a trigger consisting of a hardware stage that identifies high transverse energy particles, followed by a software stage, which applies a full event reconstruction~\cite{LHCb-DP-2012-004}. A multivariate algorithm~\cite{BBDT} is used to select candidates with secondary vertices consistent with the decay of a \bquark hadron. 

Signal $\BsToDsDs$ candidates are reconstructed in four final states: (i) $\Dsp\to\Kp\Km\pip,~\Dsm\to\Km\Kp\pim$;
(ii) $\Dsp\to\Kp\Km\pip,~\Dsm\to\pim\pip\pim$; (iii) $\Dsp\to\Kp\Km\pip, \Dsm\to\Km\pip\pim$; and
(iv) $\Dsp\to\pip\pim\pip, \Dsm\to\pim\pip\pim$. Inclusion of charge-conjugate processes, unless otherwise specified, is implicit. 
The \BzToDsD decay mode, where $\Dm\to\Kp\pim\pim$, and $\Dsp\to\Kp\Km\pip$, is used as a control channel. 
The selection requirements follow Ref.~\cite{LHCb-PAPER-2013-060}, apart from minor differences in the particle identification requirements and \BorBs candidate mass regions. 
\DorDs meson candidates are required to have masses within 25\mevcc of their known values~\cite{PDG2014} and to have a significant separation from the \BorBs vertex. 
As the signatures of $b$-hadron decays to double-charm final states are all similar, vetoes are employed to suppress the cross-feed resulting from particle misidentification, following Ref.~\cite{LHCb-PAPER-2012-050}. All \BorBs candidates are refitted, taking both \DorDs mass and vertex constraints into account~\cite{Hulsbergen:2005pu}.
A boosted decision tree (BDT)~\cite{Breiman,AdaBoost} is used to improve the signal to background ratio.
The BDT is trained with simulated decays to emulate the signal, and same-charge $\Dsp\Dsp$ and $\Dp\Dsp$ from candidates with masses on the range $5200<M(\Dsp\Dsp)<5650\mevcc$ and $5200<M(\Dp\Dsp)<5600\mevcc$, respectively. 
The selection requirement on the BDT output, which retains about 98\% of the signal events, is chosen to minimise the expected relative uncertainty in the \BsToDsDs yield.
The \BorBs candidates are required to lie in the mass regions $5300<M(\DsDs)<5450\mevcc$ for the signal and $5200<M(\DpDs)<5450\mevcc$ for the control channel, where the lower bound is chosen to suppress background contributions from \BorBs decays with excited charm mesons in the final state. The decay time distribution is fitted in the range $0.2<t<12.0\ps$ where the lower bound is chosen to reduce backgrounds from particles originating from the primary vertex.
\begin{figure}[tb]
\centering
\includegraphics[width=0.49\textwidth]{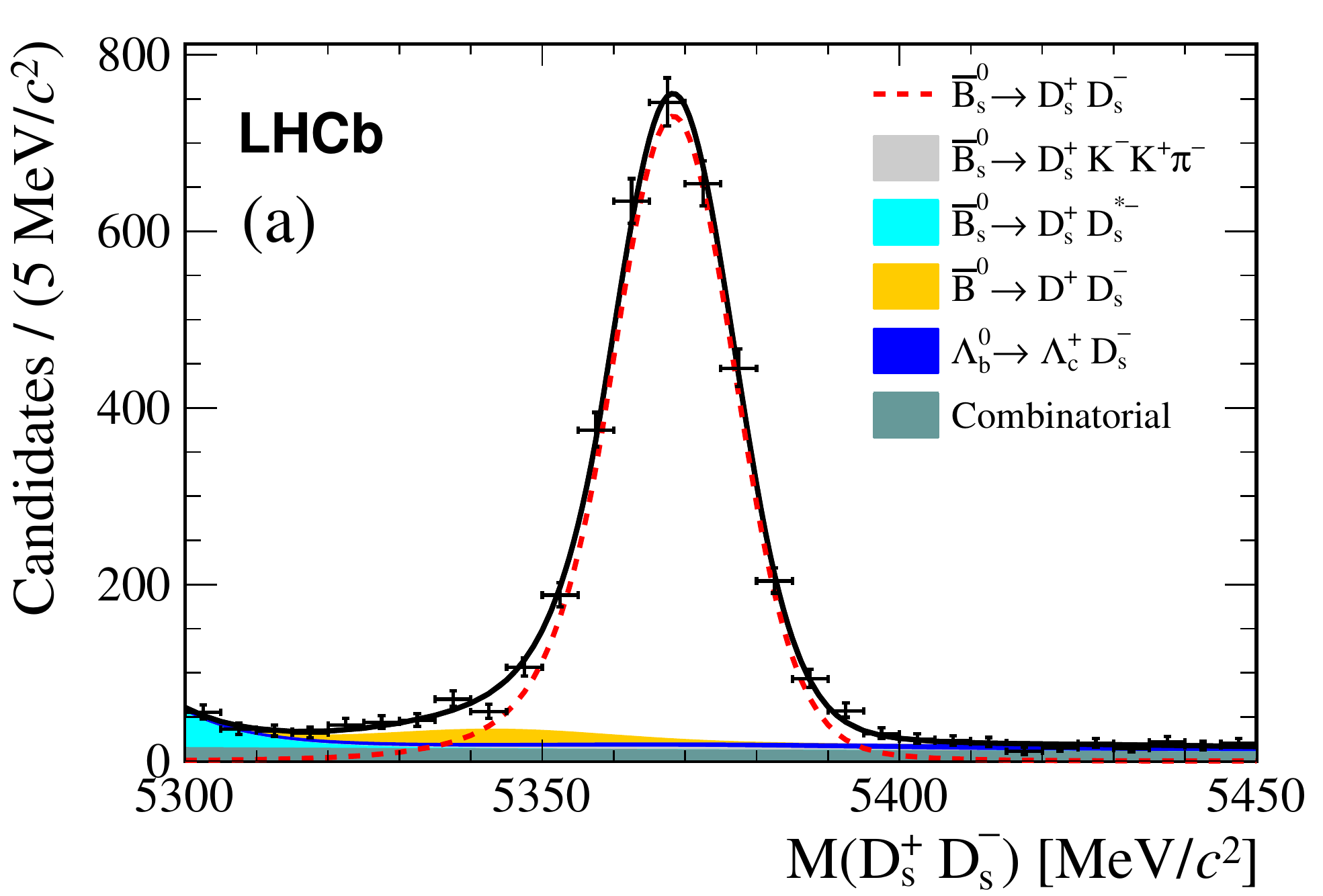}
\includegraphics[width=0.49\textwidth]{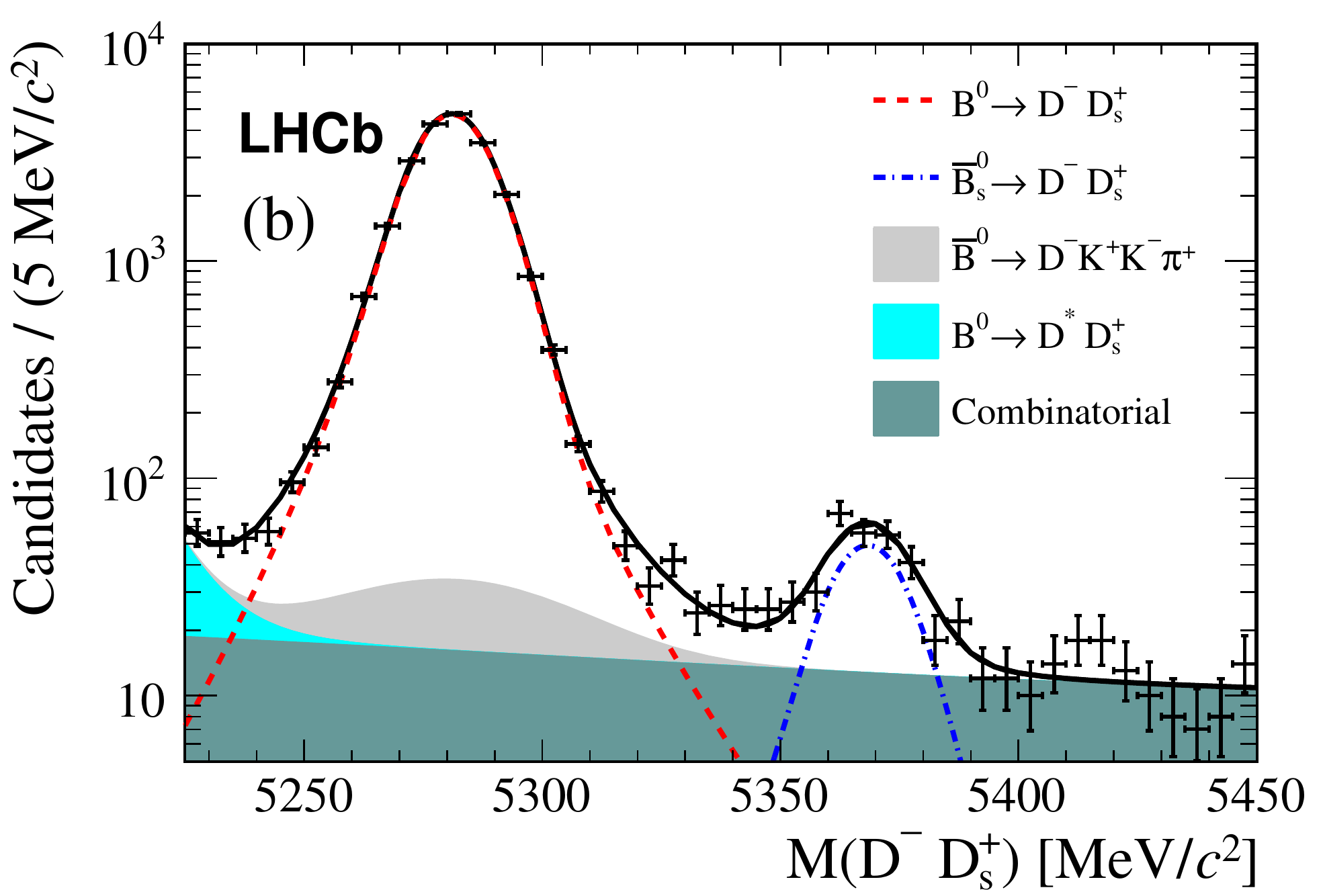}
\caption{Invariant mass distributions of (a) $\BsToDsDs$ and (b) $\BzToDsD$ candidates. The points show the data; the individual fit components are indicated in the legend; the black curve shows the overall fit.}

\label{fig:fitFullSample}
\end{figure}

The mass distributions for the signal, summed over the four final states, and the control channel
are shown in Fig.~\ref{fig:fitFullSample}, with results of unbinned maximum likelihood fits overlaid. The signal shapes
are parameterised by the sum of two asymmetric Gaussian functions with a common mean.
The background shapes are obtained from simulation~\cite{Sjostrand:2006za,*LHCb-PROC-2010-056,Lange:2001uf,Golonka:2005pn,Allison:2006ve,*Agostinelli:2002hh,*LHCb-PROC-2011-006}.
Background rates from misidentified particles are obtained from $\Dstarp\to\Dz\pip$, $\Dz\to\Km\pip$ calibration data. Signal and background components are described in Ref.~\cite{LHCb-PAPER-2013-060}.
All yields in the fits to the full data sample are allowed to vary, except that corresponding to $\bar{B}^{0}_{(s)}\to \D_{(s)}^+\Km\Kp\pim$ decays, which is fixed to be 1\% of the signal yield as determined from a fit to the $\D_{s}$ mass sidebands.
We observe $3345\pm62$ $\BsToDsDs$ signal and $21\,320\pm148$ $\BzToDsD$ control channel decays. In the \DpDs channel, we also observe a contribution from \BsToDsD as reported previously~\cite{LHCb-PAPER-2012-050}.
We use the \sPlot technique\cite{Pivk:2004ty} to obtain the decay time distribution of \BsToDsDs signal decays where the \DsDs invariant mass is the discriminating variable. 
A fit to the background-subtracted distribution of the decay time, $t$, is performed using the signal-only decay time probability density function (PDF). The negative log likelihood to be minimised is
\begin{equation}
	-\ln\mathcal{L}=-\alpha\sum_{i}^{N}W_{i}\ln \mathcal{P}(t_{i},\delta_{i},q^{\text{tag}}_{i}|\eta^{\text{tag}}_{i}),
\end{equation}
where $N$ denotes the total number of signal and background candidates in the fit region, $W_{i}$ is the signal component weight and $\alpha=\sum^{N}_{i}W_{i}/\sum^{N}_{i}W^{2}_{i}$~\cite{2009arXiv0905.0724X}. The invariant mass is not correlated with the reconstructed decay time or its uncertainty, nor with flavour tagging output, for signal and background. 
The signal PDF, $\mathcal{P}$, includes detector resolution and acceptance effects and requires knowledge of the \Bs (\Bsb) flavour at production,
\begin{equation}
	\mathcal{P}(t,\delta,q^{\text{tag}}|\eta^{\text{tag}}) =  R(\hat{t},q^{\text{tag}}|\eta^{\text{tag}}) \otimes G(t-\hat{t}|\delta) \times \epsilon^{\Dsp\Dsm}_{\text{data}}(t),
\end{equation}
where $\hat{t}$ is the decay time in the absence of resolution effects, $R(\hat{t},q^{\text{tag}}|\eta^{\text{tag}})$ describes the rate including imperfect knowledge of the initial $\BorBbar^0_{s}$ flavour through the flavour tag, $q^{\text{tag}}$, and the wrong-tag probability estimate $\eta^{\text{tag}}$. The flavour tag, $q^{\text{tag}}$, is $-1$ for \Bsb, $+1$ for \Bs and zero for untagged candidates.
The calibrated decay time resolution is $G(t-\hat{t}|\delta)$ where $\delta$ is the decay time error estimate, and $\epsilon^{\Dsp\Dsm}_{\text{data}}(t)$ is the decay time acceptance.

Allowing for \CP violation in decay, the decay rates of $\BorBbar^0_{s}$ mesons ignoring detector effects can be written as
\begin{align}
        \Gamma(\hat{t}) = {\cal N} e^{- \Gs \hat{t}}  & \left [\cosh\left(\frac{\DGs}{2}\hat{t}\right) -\frac{2|\lambda|\cos\phis}{1+|\lambda|^{2}}\sinh\left(\frac{\DGs}{2}\hat{t}\right) \right . \nonumber \\
                                                                                                                                                                                    & \left . +\frac{1-|\lambda|^{2}}{1+|\lambda|^{2}}\cos(\dms \hat{t}) -\frac{2|\lambda|\sin\phis}{1+|\lambda|^{2}}\sin(\dms \hat{t}) \right ], \label{eq:CPrateeven}\\ 
        \bar{\Gamma}(\hat{t}) = \left \vert \frac{p}{q}\right \vert^{2} {\cal N} e^{- \Gs \hat{t}}  & \left [\cosh\left(\frac{\DGs}{2}\hat{t}\right) -\frac{2|\lambda|\cos\phis}{1+|\lambda|^{2}}\sinh\left(\frac{\DGs}{2}\hat{t}\right) \right . \nonumber \\
                                                                                                                                                                                                                                                                                                               & \left . -\frac{1-|\lambda|^{2}}{1+|\lambda|^{2}}\cos(\dms \hat{t}) + \frac{2|\lambda|\sin\phis}{1+|\lambda|^{2}}\sin(\dms \hat{t}) \right ], \label{eq:CPrateevenbar}
\end{align}
where $\Gs\equiv(\GL + \GH)/2$ is the average decay width of the light and heavy mass eigenstates, $\DGs\equiv\GL-\GH$ is their decay width difference and $\dms\equiv\mH-\mL$ is their mass difference. As \dms is large~\cite{LHCb-PAPER-2013-006} and the production asymmetry is small~\cite{LHCb-PAPER-2014-042}, the effect of the production asymmetry is negligible and so the constant $\cal{N}$ is the same for both \Bs and \Bsb mesons. Similarly we do not consider a tagging asymmetry in the fit as this is known to be consistent with zero. \CP violation in mixing and decay is parameterised by the factor $\lambda \equiv \frac{q}{p}\frac{\bar{A}_{f}}{A_{f}}$, with $\phis\equiv-\arg(\lambda)$. The terms ${A}_{f}$ ($\bar{A}_{f}$) are the amplitudes for the $\Bs$ ($\Bsb$) decay to the final state $f$, which in this case is $f=\DsDs$, and the complex parameters $ p=\langle \Bs | \B_{L}\rangle $ and  $q=\langle \Bsb | \B_{L} \rangle $ relate the mass and flavour eigenstates.
The factor $|p/q|^{2}$ in Eq.~(\ref{eq:CPrateevenbar}) is related to the flavour-specific \CP asymmetry, $a^{s}_{\text{sl}}$, by
\begin{equation}
                a^{s}_{\text{sl}} = \frac{\vert p/q\vert^{2} - \vert q/p\vert^{2}}{\vert p/q\vert^{2} + \vert q/p\vert^{2}} \approx \vert p/q\vert^{2} - 1. 
        \end{equation} 
LHCb has measured $a^{s}_{\text{sl}}=(-0.06\pm0.50\stat\pm0.36\syst)\%$ \cite{LHCB-PAPER-2013-033}, implying $\vert p/q\vert^{2}=0.9994\pm0.0062$. We assume that it is unity in this analysis and that any observed deviation of $|\lambda|$ from 1 is due to \CP violation in the decay, \ie $|\bar{A}_{f}/A_{f}|\neq1$.

The initial flavor of the signal b hadron is determined using two methods. In hadron collisions, \bquark hadrons are mostly produced as pairs: the opposite-side (OS) tagger~\cite{LHCb-PAPER-2011-027} determines the flavour of the other \bquark hadron in the event by identifying the charges of the leptons and kaons into which it decays, or the net charge of particles forming a detached vertex consistent with that of a \bquark hadron. The neural network same-side (SS) kaon tagger~\cite{LHCb-PAPER-2013-002} exploits the hadronisation process in which the fragmentation of a $\bar{\bquark}(\bquark)$ into a $\Bs(\Bsb)$ meson leads to an extra $\bar{\squark}(\squark)$ quark, which often forms a $\Kp(\Km)$ meson, the charge of which identifies the initial $\BorBbar^0_{s}$ flavour. The SS kaon tagger uses an improved algorithm with respect to Ref.~\cite{LHCb-PAPER-2013-002} that enhances the fraction of correctly tagged mesons by 40\%. In both tagging algorithms a per-event wrong-tag probability estimate, $\eta^{\text{tag}}$, is determined, based on the output of a neural network trained on either simulated \BsToDsPi events for the SS tagger, or, in the case of the OS algorithm, using a data sample of \BToJpsiK decays. The taggers are then calibrated in data using flavour-specific decay modes in order to provide a per-event wrong-tag probability, $\brabar{\omega}(\eta^{\text{tag}})$, for an initial flavour $\BorBbar^0_{s}$ meson. The calibration is performed separately for the two tagging algorithms, which are then combined in the fit. The effective tagging power is parameterised by $\effD$ where $D\equiv(1-2\mistag)$ and \etag is the fraction events tagged by the algorithm.

The combined effective tagging power is $\effD=(5.33\pm0.18\stat\pm0.17\syst)\%$, comparable to that of other recent analyses~\cite{LHCB-PAPER-2014-038}.
The rate expression including flavour tagging is
\begin{eqnarray}
	R(\hat{t},q^{\text{OS}}|\eta^{\text{OS}}, q^{\text{SS}}|\eta^{\text{SS}}) = & (1 + q^{\text{OS}}[1 - 2\omega^{\text{OS}}])(1 + q^{\text{SS}}[1 - 2\omega^{\text{SS}}])\Gamma(\hat{t}) + \nonumber \\
																																				      & (1 -  q^{\text{OS}}[1 - 2\bar{\omega}^{\text{OS}}])(1 - q^{\text{SS}}[1 - 2\bar{\omega}^{\text{SS}}])\bar{\Gamma}(\hat{t}) .\phantom{+}
\end{eqnarray}

\begin{figure}
        \includegraphics[width=0.45\textwidth]{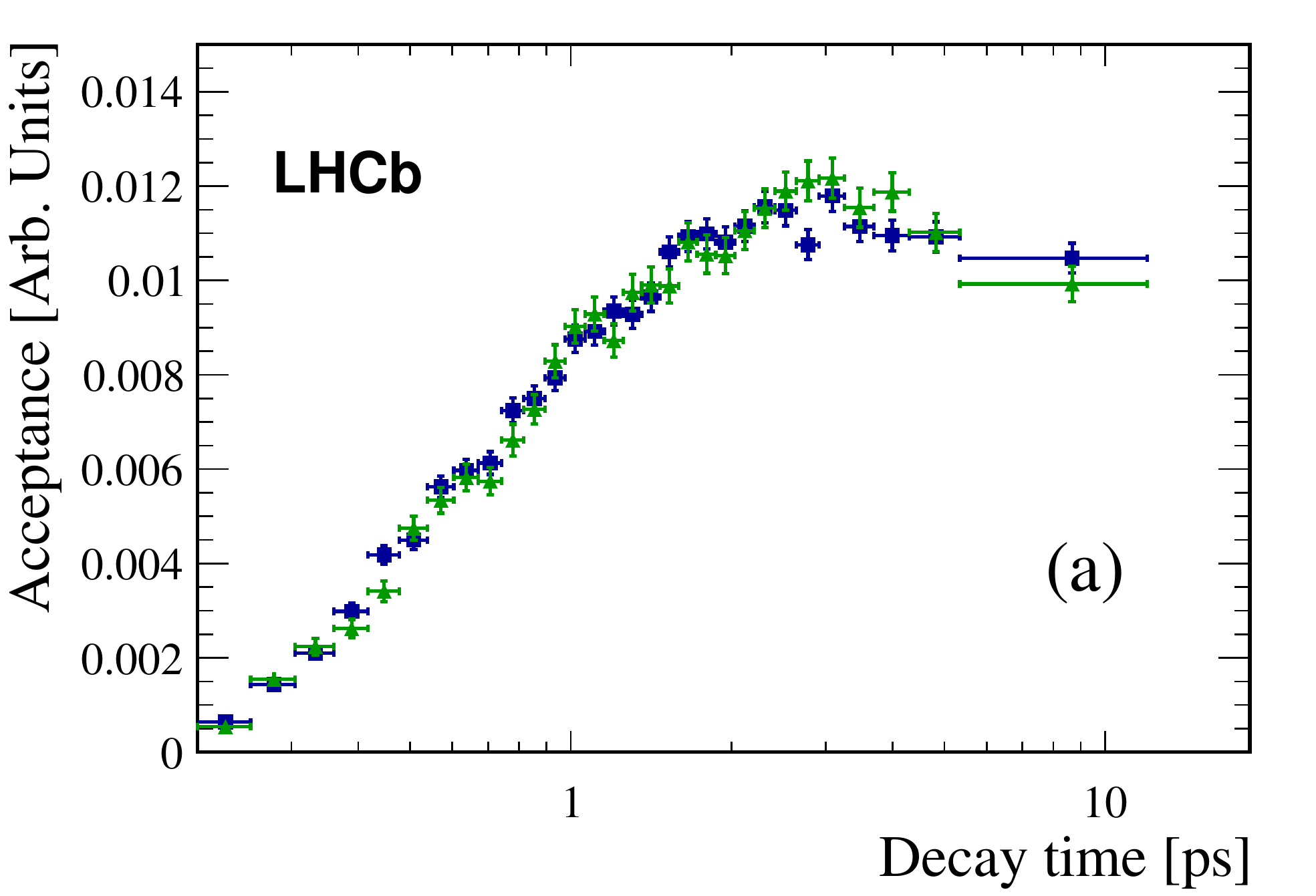}
        \includegraphics[width=0.45\textwidth]{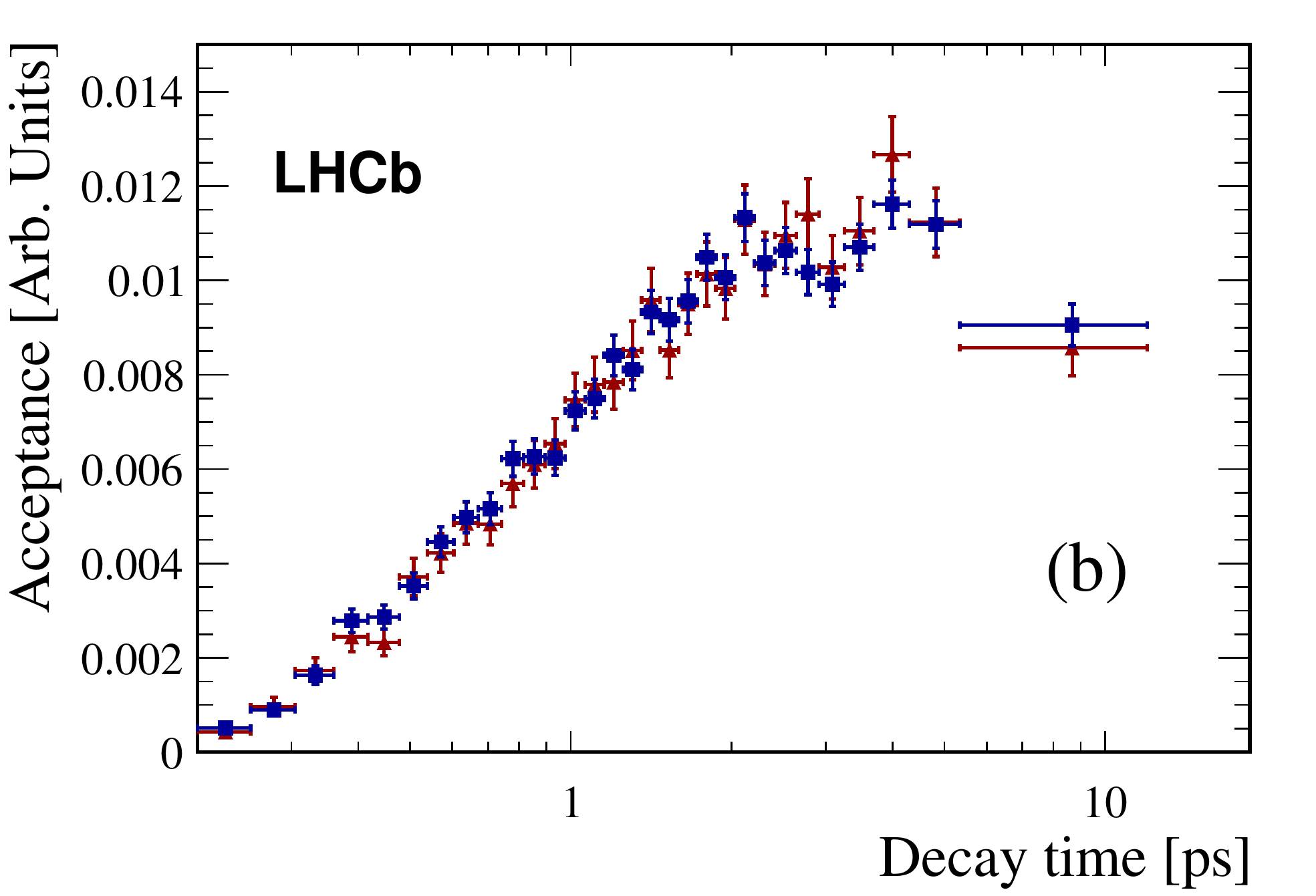}
        \caption{\label{fig:acceptance} Decay time acceptances in simulation and data: (a) the \BzToDsD acceptance in data (green triangles) and simulation (blue squares), (b) the \BsToDsDs acceptance in simulation (blue squares) and the \BzToDsD acceptance corrected for \BsToDsDs (red triangles). The correction is described in detail in the text.}
\end{figure}

The track reconstruction, trigger and selection efficiencies vary as a function of decay time, requiring that an acceptance function is included in the fit. The \BsToDsDs acceptance is determined using
\begin{equation}
        \varepsilon^{\DsDs}_{\text{data}}(t) = \varepsilon^{\DpDs}_{\text{data}}(t) \times \frac{\varepsilon^{\DsDs}_{\text{sim}}}{\varepsilon^{\DpDs}_{\text{sim}}}(t), \label{eq:acceptance}
\end{equation}
where $\varepsilon^{\DpDs}_{\text{data}}(t)$ is the efficiency associated with the \BzToDsD control channel as determined directly from the data and $\varepsilon^{\DsDs}_{\text{sim}}/\varepsilon^{\DpDs}_{\text{sim}}(t)$ is the relative efficiency obtained from simulation after all selections are applied. This correction accounts for the differences in lifetime as well as small kinematic differences between the signal and control channels. The first factor in Eq.~(\ref{eq:acceptance}) is
\begin{equation}
        \varepsilon^{\DpDs}_{\text{data}}(t) = \frac{N^{\DpDs}_{\text{data}}(t)}{\mathcal{N}e^{-\Gamma_{d}\hat{t}}\otimes G(t-\hat{t}|\sigma_{\text{eff}})},
\end{equation} 
where $N^{\DpDs}_{\text{data}}(t)$ denotes the number of \BzToDsD signal decays in a given bin of the decay time distribution, $\mathcal{N}e^{-\Gamma_{d}\hat{t}}$ is an exponential with decay width equal to that of the world average value for \Bd mesons~\cite{PDG2014}, $\cal{N}$ is a constant and $G(t-\hat{t}|\sigma_{\text{eff}})$ is a Gaussian resolution function with width $\sigma_{\text{eff}}=54\fs$, determined from simulation.  
In the fit, the acceptance is implemented as a histogram. The binning scheme is chosen to maintain approximately equal statistical power in each bin. Figure~\ref{fig:acceptance}(a) shows $\varepsilon^{\DpDs}_{\text{data}}(t)$ and $\varepsilon^{\DpDs}_{\text{sim}}(t)$, while  Fig.~\ref{fig:acceptance}(b) shows $\varepsilon^{\DsDs}_{\text{sim}}(t)$ and $\varepsilon^{\DsDs}_{\text{data}}(t)$ as used in the fit to extract \phis. The procedure is verified by fitting for the decay width in both the signal and the control channels, where the results are found to be consistent with the published values.

The fit to determine \phis uses a decay time uncertainty estimated in each event and obtained from the constrained vertex fit from which the decay time is determined. The resolution function is
\begin{equation}
	G(t-\hat{t}|\delta) = \frac{1}{\sqrt{2\pi}\sigma(\delta)} e^{-\frac{1}{2}\left  ( \frac{t-\hat{t}}{\sigma(\delta)}\right )^{2}}. 
\end{equation}
The per-event resolution, $\sigma(\delta)$, is calibrated using simulated signal decays by fitting the effective resolution, $\sigma_{\text{eff}}$, in bins of the per-event decay time error estimate, $\sigma_{\text{eff}} = q_{0} + q_{1}\delta$. The effective resolution is determined by fitting to the event-by-event decay time difference between the reconstructed and generated decay time in simulated signal decays. The effective resolution is the sum in quadrature of the widths of two Gaussian functions contributing with their corresponding fractions. The values $q_{0}=8.9\pm1.3\fs$ and $q_{1}=1.014\pm0.036$ are obtained from the fit, resulting in a calibrated effective resolution of 54\fs.

\begin{figure}\centering
        \includegraphics[width=0.7\textwidth]{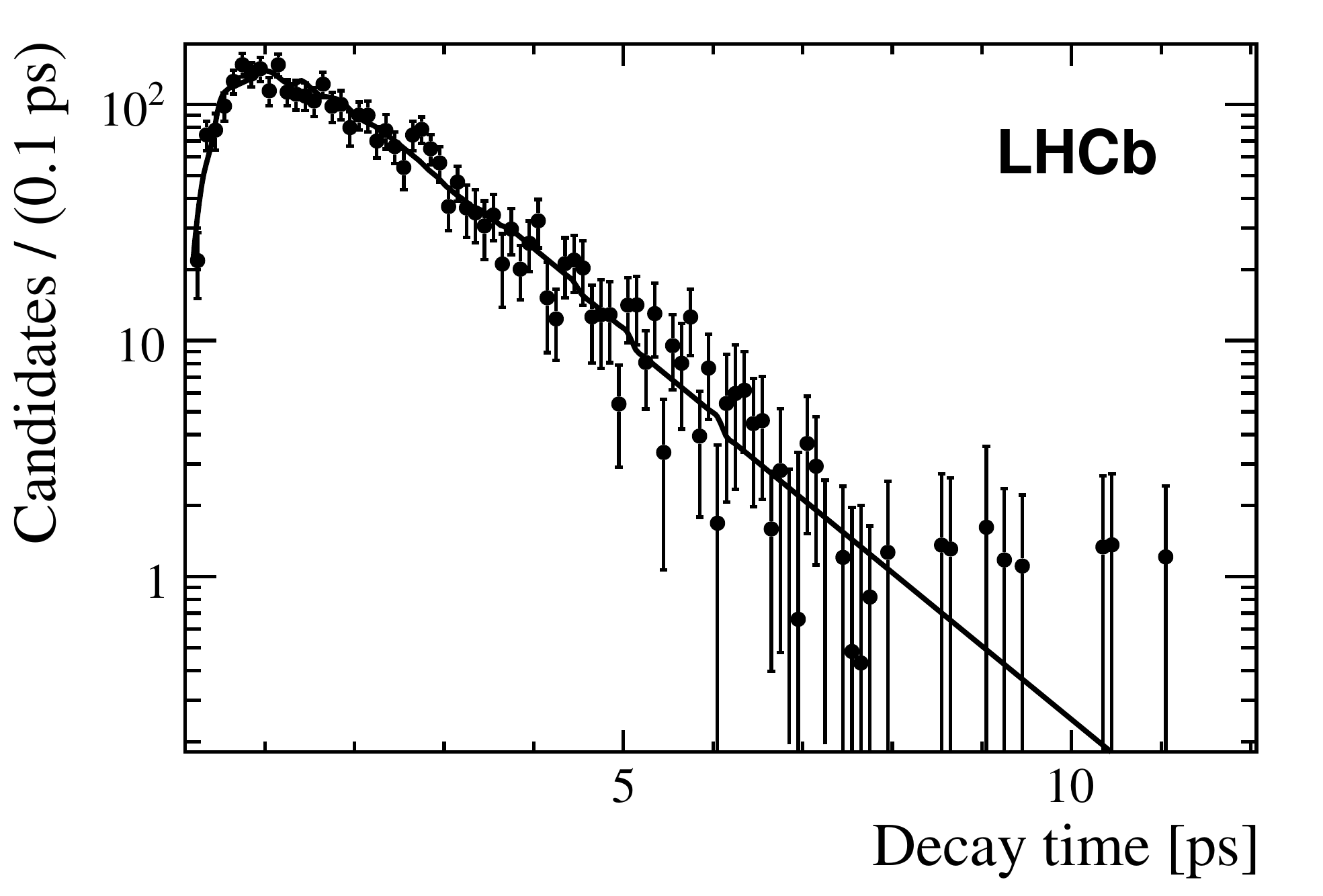}
        \caption{\label{fig:timefit}Distribution of the decay time for \BsToDsDs signal decays with background subtracted using the \sPlot method, along with the fit as described in the text. Discontinuities in the fit line shape are a result of the binned acceptance.}
\end{figure}

In the fits that determine \phis, we apply Gaussian constraints to the average decay width, $\Gs=0.661\pm0.007\invps$, the decay width difference, $\DGs=0.106\pm0.013\invps$~\cite{LHCb-PAPER-2013-002}, the mixing frequency, $\dms=17.168\pm0.024\invps$~\cite{LHCb-PAPER-2013-006} and the flavour tagging and resolution calibration parameters. The correlation between \Gs and \DGs is accounted for in the fit. 
Two fits to the data are performed, one assuming no \CP violation in decay, \ie $|\lambda|=1$, and a second where this assumption is removed. 
The fit is validated using pseudoexperiments and simulated LHCb events. 

\begin{table}\centering
        \caption{\label{tab:systs} Summary of systematic uncertainties not already accounted for in the fit, where $\sigma$ denotes the statistical uncertainty.}
        \begin{tabular}{lllll}
		Systematic uncertainty    & \multicolumn{1}{c}{$\phis$ ($|\lambda|=1$)} && \multicolumn{1}{c}{$\phis$}            & \multicolumn{1}{c}{$|\lambda|$}       \\ \hline
		Resolution & $\pm0.098~\sigma$       && $\pm0.094~\sigma$  & $\pm0.100~\sigma$ \\ 
		Mass & $\pm0.044~\sigma$       && $\pm0.043~\sigma$  & $\pm0.010~\sigma$ \\
		Acceptance (model) & $\pm0.022~\sigma$       && $\pm0.027~\sigma$  & $\pm0.027~\sigma$ \\
		Acceptance (stat.) & $\pm0.013~\sigma$       && $\pm0.013~\sigma$  & $\pm0.014~\sigma$ \\
		Background subtraction      & $\pm0.009~\sigma$      && $\pm0.008~\sigma$ & $\pm0.046~\sigma$ \\ \hline
		Total                   & $\pm0.11~\sigma$       && $\pm0.11~\sigma$  & $\pm0.11~\sigma$ \\
        \end{tabular}
\end{table}
The systematic uncertainties on $\phis$ and $|\lambda|$ that are not accounted for by the use of Gaussian constraints are summarised in Table~\ref{tab:systs}. 
The systematic uncertainty associated with the resolution calibration in simulated events is studied by generating pseudoexperiments with an alternative resolution parameterisation ($q_{0}=0$, $q_{1}\in[1.25,1.45]$~\cite{LHCb-PAPER-2013-006}) obtained in \Bsb decays in data. 
The effect of mismodelling of the mass PDF is studied by fitting using a larger mass window and including an additional background component from $\BsToDsstDsst$. 
The effect of mismodelling the acceptance distribution is studied by fitting the \BsToDsD derived acceptance in pseudoexperiments generated with the acceptance distribution determined entirely from \BsToDsDs simulation. 
The uncertainty due to the finite size of the simulated data samples used to determine the acceptance correction is evaluated by fitting to the data 500 times with Gaussian fluctuations around the bin values with a width equal to the statistical uncertainties. 
We evaluate the uncertainty due to the use of the \sPlot method for background subtraction by fitting to simulated events, once with only signal candidates, and again to the \sPlot determined from a mass fit to a sample containing the signal and background in proportions determined from data. 

Assuming no \CP violation in decay, we find
\begin{equation*}
        \phis=0.02\pm0.17\stat\pm0.02\syst\rad,
\end{equation*} 
where the first uncertainty is statistical and the second is systematic. 
In a fit to the same data in which we allow for the presence of \CP violation in decay we find
\begin{equation*}
        \phis=0.02\pm0.17\stat\pm0.02\syst\rad, \qquad |\lambda| = 0.91~^{+0.18}_{-0.15}\stat\pm0.02\syst,
\end{equation*}
where \phis and $|\lambda|$ have a correlation coefficient of $3\%$.  This measurement is consistent with no \CP violation. The decay time distribution and the corresponding fit projection for the case where \CP violation in decay is allowed are shown in Fig.~\ref{fig:timefit}. 

In conclusion, we present the first analysis of the time evolution of flavour-tagged \BsToDsDs decays.
We measure the \CP-violating weak phase $\phis$, allowing for the presence of \CP violation in decay, and find that it is consistent with the Standard Model expectation and with measurements of \phis in other decay modes.

\section*{Acknowledgements}
\noindent We express our gratitude to our colleagues in the CERN
accelerator departments for the excellent performance of the LHC. We
thank the technical and administrative staff at the LHCb
institutes. We acknowledge support from CERN and from the national
agencies: CAPES, CNPq, FAPERJ and FINEP (Brazil); NSFC (China);
CNRS/IN2P3 (France); BMBF, DFG, HGF and MPG (Germany); SFI (Ireland); INFN (Italy); 
FOM and NWO (The Netherlands); MNiSW and NCN (Poland); MEN/IFA (Romania); 
MinES and FANO (Russia); MinECo (Spain); SNSF and SER (Switzerland); 
NASU (Ukraine); STFC (United Kingdom); NSF (USA).
The Tier1 computing centres are supported by IN2P3 (France), KIT and BMBF 
(Germany), INFN (Italy), NWO and SURF (The Netherlands), PIC (Spain), GridPP 
(United Kingdom).
We are indebted to the communities behind the multiple open 
source software packages on which we depend. We are also thankful for the 
computing resources and the access to software R\&D tools provided by Yandex LLC (Russia).
Individual groups or members have received support from 
EPLANET, Marie Sk\l{}odowska-Curie Actions and ERC (European Union), 
Conseil g\'{e}n\'{e}ral de Haute-Savoie, Labex ENIGMASS and OCEVU, 
R\'{e}gion Auvergne (France), RFBR (Russia), XuntaGal and GENCAT (Spain), Royal Society and Royal
Commission for the Exhibition of 1851 (United Kingdom).

\clearpage
\addcontentsline{toc}{section}{References}
\setboolean{inbibliography}{true}
\ifx\mcitethebibliography\mciteundefinedmacro
\PackageError{LHCb.bst}{mciteplus.sty has not been loaded}
{This bibstyle requires the use of the mciteplus package.}\fi
\providecommand{\href}[2]{#2}

\newpage
\newpage
\centerline{\large\bf LHCb collaboration}
\begin{flushleft}
\small
R.~Aaij$^{41}$, 
C.~Abell\'{a}n~Beteta$^{40}$, 
B.~Adeva$^{37}$, 
M.~Adinolfi$^{46}$, 
A.~Affolder$^{52}$, 
Z.~Ajaltouni$^{5}$, 
S.~Akar$^{6}$, 
J.~Albrecht$^{9}$, 
F.~Alessio$^{38}$, 
M.~Alexander$^{51}$
% ...

{\footnotesize \it
% ...
$ ^{5}$Clermont Universit\'{e}, Universit\'{e} Blaise Pascal, CNRS/IN2P3, LPC, Clermont-Ferrand, France\\
$ ^{6}$CPPM, Aix-Marseille Universit\'{e}, CNRS/IN2P3, Marseille, France\\
% ...
$ ^{9}$Fakult\"{a}t Physik, Technische Universit\"{a}t Dortmund, Dortmund, Germany\\
% ...
$ ^{37}$Universidad de Santiago de Compostela, Santiago de Compostela, Spain\\
$ ^{38}$European Organization for Nuclear Research (CERN), Geneva, Switzerland\\
% ...
$ ^{40}$Physik-Institut, Universit\"{a}t Z\"{u}rich, Z\"{u}rich, Switzerland\\
$ ^{41}$Nikhef National Institute for Subatomic Physics, Amsterdam, The Netherlands\\
% ...
$ ^{46}$H.H. Wills Physics Laboratory, University of Bristol, Bristol, United Kingdom\\
% ...
$ ^{51}$School of Physics and Astronomy, University of Glasgow, Glasgow, United Kingdom\\
$ ^{52}$Oliver Lodge Laboratory, University of Liverpool, Liverpool, United Kingdom\\
% ...
}
\end{flushleft}

\end{document}